\documentclass[CEJP,PDF]{cej} 
\usepackage{layout}
\usepackage{amsmath}
\usepackage{textcomp}

\title{The imprints of superstatistics in multiparticle production processes}

\articletype{Research Article}

\author{Grzegorz Wilk\inst{1}\email{wilk@fuw.edu.pl},
        Zbigniew W\l odarczyk\inst{2}\email{zbigniew.wlodarczyk@ujk.edu.pl}}

\institute{\inst{1} National Centre for Nuclear Research, Theoretical Physics Department,\\
                    Ho\.za 69, 00-681 Warsaw, Poland
           \inst{2} Institute of Physics, Jan Kochanowski
                    University,\\
                    \'Swi\c{e}tokrzyska 15, 25-405 Kielce, Poland
            }

\abstract{We provide an update of the overview of imprints of
Tsallis nonextensive statistics seen in a multiparticle production
processes. They reveal an ubiquitous presence of power law
distributions of different variables characterized by the
nonextensivity parameter $q > 1$. In nuclear collisions one
additionally observes a $q$-dependence of the multiplicity
fluctuations reflecting the finiteness of the hadronizing source.
We present sum rules connecting parameters $q$ obtained from an
analysis of different observables, which allows us to combine
different kinds of fluctuations seen in the data and analyze an
ensemble in which the energy (E), temperature (T) and multiplicity
(N) can all fluctuate. This results in a generalization of the so
called Lindhard's thermodynamic uncertainty relation. Finally,
based on the example of nucleus-nucleus collisions (treated as a
quasi-superposition of nucleon-nucleon collisions) we demonstrate
that, for the standard Tsallis entropy with degree of
nonextensivity $q < 1$, the corresponding standard Tsallis
distribution is described by $q' = 2 - q > 1$.}


\begin{document}
\maketitle


\section{\label{sec:I}Introduction}

Multiparticle production processes are the main source of our
knowledge of the properties of matter formed in extreme
conditions. In fact, they are sometimes regarded as an
illustration, in laboratory conditions, of the state apparently
existing just after the Big Bang. From the very beginning for
their investigation it was obvious that the best way of their
description and understanding is the approach of statistical
mechanics\footnote{See, for example \cite{MG_rev} and references
therein.}. For a decade now we advocate (see review \cite{WW_epja}
for references) that the quality of data is high enough to observe
the deviations from the usual statistical approach based on a
Boltzman-Gibbs ensemble towards a more refined one based on its
generalization, for example given by Tsallis statistics
\cite{T,T2,T3}\footnote{For an updated bibliography on this
subject see http://tsallis.cat.cbpf.br/biblio.htm.}. From the very
beginning, we have identified the $q$ parameter occurring in
Tsallis statistics with the measure of some intrinsic fluctuations
present in the system \cite{qWW,qWW1}. They are natural in systems
where the heat bath is not homogeneous. In short: if the scale
parameter $\tilde{T}$ ("temperature") in the usual Boltzman
distribution, $f(E) \propto \exp\left( - E/{\tilde{T}}\right)$,
fluctuates according to a gamma distribution, $g(1/{\tilde{T}})
\propto \left\{ T/\left[(q -
1)\tilde{T}\right]\right\}^{(2-q)/(q-1)} \exp\left\{ - T/\left[(q
- 1)\tilde{T}\right]\right\}$, with fluctuations given by $q = 1 +
Var(1/\tilde{T})/\langle 1/\tilde{T}\rangle^2$, as result one gets
the power-like Tsallis distribution:
\begin{eqnarray}
h_q(E) = \frac{2-q}{T}\exp_q \left(-\frac{E}{T}\right) =
\frac{2-q}{T}\left[1 - (1-q)\frac{E}{T}\right]^{\frac{1}{1-q}}
\label{eq:Tsallis}.
\end{eqnarray}
This was quickly recognized as an example of so called {\it
superstatistics} \cite{sS,sS1} and, in what follows, we shall
understand it in this way, i.e., we shall mainly concentrate on
the notion of fluctuations represented by the parameter $q$.

The conclusion of \cite{qWW,qWW1} was reinforced by a more refined
analysis in \cite{BJ,BJ1} and generalized in \cite{WW_epja} by
additionally allowing for the heat bath to exchange energy with
its surroundings. One then gets the same Tsallis distribution as
before, but with a $q$-dependent effective temperature, $T_{eff} =
T + (q - 1)T_{*}$, replacing $T$ in Eq. (\ref{eq:Tsallis}). Here
$T_{*}$ is a new parameter depending on the transport properties
of the space surrounding the emission region. In \cite{WW_epja},
where this concept was first introduced for heavy ion collisions,
$T_{*} = \phi /(Dc_{P}\rho)$, $\phi $ is the energy transfer
taking place between the source and its surroundings\footnote{It
can be connected with viscosity $\eta$ by $\phi = \eta f(u)$,
where $f(u) = \left(\partial u_i/\partial x_k + \partial
u_k/\partial x_l\right)^2$ ($u$ being velocity); $D$, $c_{P}$ and
$\rho$ are, respectively, the strength of the temperature
fluctuations, the specific heat under constant pressure and the
density.}. This quantity is supposed to model the possible
transfer of energy from the central region of nucleus-nucleus
interaction towards the spectators not participating in the
collision (when $T_{*} < 0$)\footnote{It is interesting to note
that in cosmic ray physics it can be argued \cite{qCR} that the
corresponding $T_{*} > 0$ and describes the transfer of energy
from the surroundings to cosmic ray particles accelerated in outer
space.}.

For nuclear collisions the possible transfer of energy between the
region of interaction and the spectators depends on its size (or
on centrality of the reaction). Accounting for it thus explains
automatically the observed $q$-dependence of the multiplicity
fluctuations measured in nuclear collisions on the centrality of
collision \cite{WWprc} (cf. also \cite{qcompilation}). Assuming
that the size of the thermal system produced in heavy ion
collisions is proportional to the number of nucleons participating
in collision, $N_P$, one gets that ($C_V$ is heat capacity under
constant volume)
\begin{equation}
 q - 1 = \frac{1}{aN_P}\left( 1 - \frac{N_P}{A}\right),\qquad a =
 \frac{C_V}{N_P}  \label{eq:qN_P}
\end{equation}
which nicely fits the data \cite{WWprc}.

Since the review \cite{WW_epja}, Tsallis distributions have been
used more widely, for example, they were successfully applied to
the analysis of the recent PHENIX \cite{PHENIX} and LHC CMS
\cite{LHC_CMS,LHC_CMS1} data (cf. also compilation
\cite{qcompilation}). In what follows we present our recent
results in this field in more detail: a derivation of $q$-sum
rules unifying different fluctuations - in Section
\ref{section:II}, a generalization of thermodynamic uncertainty
relations and their applications - in Section \ref{section:III}.
Section \ref{section:IV} contains our present result, namely,
using as an example the experimental data on nucleus-nucleus
collisions (treated as quasi-superposition of nucleon-nucleon
collisions), we demonstrate the interrelation between
nonextensivity parameters obtained from Tsallis entropy and
Tsallis distributions. Section \ref{section:V} summarizes our
work.

\section{\label{section:II}$q$-sum rules}

As argued above, the parameter $q$ provides a useful measure of
intrinsic (nonstatistical) fluctuations in the system \cite{qWW}.
In multiparticle production processes the particles produced are
characterized by their positions in phase space, in particular the
measured momentum $\vec{p}$ is decomposed into longitudinal
component (along the direction of colliding particles), $p_L = m_T
\sinh y$, and transverse component, $p_T$, with $p =
\sqrt{|\vec{p}^2|} = \sqrt{p^2_L + p_T^2}$ (here $m_T = \sqrt{m^2
+ p_T^2}$ denotes the so called "transverse mass" of the particle
and $y = \frac{1}{2}\ln\frac{E + p_L}{E - p_L}$ its "rapidity";
$E$ is the energy of the particle and $m$ is its mass). Because in
most cases data are presented in the form of distributions in
rapidity $y$ (i.e., they are integrated over $p_T$), $dN/dy$, and
as distributions in $p_T$ (i.e., they are integrated over $y$),
$dN/dp_T$, one is confronted with two different fluctuations: in
longitudinal phase space, characterized by $q = q_L$, and in
transverse phase space characterized by $q=q_T$. It turns out that
the strengths of both fluctuations measured by $q$ is different,
whereas $q_L - 1 \sim 0.1 - 0.3$ and grows with energy of
collision (measured mainly in $pp$ and ${\bar p}p$ collisions),
transverse fluctuations are much weaker, $q_T - 1 \sim 0.01-0.1$,
vary slowly with energy and depend slightly on whether one
observes elementary collisions or collisions between nuclei
\cite{QlQt2,QlQt3,MB1,MB2}.

There is another source of knowledge concerning fluctuations,
namely observed multiplicity distributions, $P(N)$. It turns out
that temperature fluctuations in the form of a gamma distribution
leading to Eq. (\ref{eq:Tsallis}) result in substantial broadening
of the corresponding multiplicity distributions. This changes from
poissonian form characteristic for exponential distributions,
$P(N) =  \bar{N}^N \exp\left ( - \bar{N}\right)/N!$ (where
$\bar{N} = E/T$), to negative binomial form (NB) for Tsallis
distributions, Eq. (\ref{eq:Tsallis}) (cf., \cite{fluct}, for
details)\footnote{Notice that in the limiting cases of
$q\rightarrow 1$ one has $k\rightarrow \infty$ and (\ref{eq:NBD})
becomes a poissonian distribution, whereas for $q\rightarrow 2$ on
has $k\rightarrow 1$ and (\ref{eq:NBD}) becomes a geometrical
distribution.},
\begin{equation}
P(N)\, =\, \frac{\Gamma(N+k)}{\Gamma(N+1)\Gamma(k)}\frac{\left(
\frac{\langle N\rangle}{k}\right)^N}{\left( 1 + \frac{\langle
N\rangle}{k}\right)^{(N+k)}};\quad {\rm where}\quad
k=\frac{1}{|q-1|}.\label{eq:NBD}
\end{equation}
The nonextensivity parameter $q$ reflects here fluctuations {\it
in the whole of phase space} and can be (and usually is),
different from the previously obtained $q_L$ and $q_T$. In
\cite{QlQt2,QlQt3} it was proposed that because $q-1 =
\sigma^2(T)/T^2$ (i.e., it is given by fluctuations of total
temperature $T$), then assuming that $\sigma^2(T) = \sigma^2(T_L)
+ \sigma^2(T_T)$, the resulting values of $q$ should not be too
different from
\begin{equation}
q\, =\, \frac{q_L\, T_L^2\, +\, q_T\, T^2_T}{T^2}\, -\,
        \frac{T^2_L\, +\, T^2_T}{T^2}\, +\, 1  .
\label{eq:qqq}
\end{equation}
Therefore for the observed dominance of longitudinal (partition)
temperature over the transverse one, $T_L \gg T_T$, one should
expect that $q \sim q_L$, which is indeed observed
\cite{QlQt2,QlQt3}. This is the first sum rule observed for
parameters $q$ obtained from different measurements.

In cases where other variables in addition to $T$ also fluctuate
(usually fluctuations of temperature are deduced either from data
averaged over other fluctuations or from data accounting also for
fluctuations of other variables) one should refine the
experimentally evaluated $q$. In this case, when extracting $q$
from distributions of $dN/dy$, one finds that (cf., \cite{WWW} for
details)
\begin{equation}
q - 1 \stackrel{def}{=} \frac{Var(T)}{\langle T\rangle^2} =
\frac{Var(z)}{\langle z\rangle^2} - \frac{Var\left(
m_T\right)}{\langle m_T \rangle^2};  \qquad z = \frac{m_T}{T}.
\label{eq:sumrule}
\end{equation}
This is the second sum rule for parameters $q$ obtained from
different measurements. It connects the total $q$, which can be
obtained from an analysis of the NB form of the measured
multiplicity distributions, $P(N)$, with $q_L -1 = Var(z)/\langle
z\rangle^2$, obtained from fitting rapidity distributions and
$Var\left(m_T\right)/\langle m_T\rangle^2$ obtained from data on
transverse mass distributions. When extracting $q$ from
distributions of $dN/dm_T$, we proceed analogously with $z=\cosh
y/T$.

\section{\label{section:III}Generalized thermodynamic uncertainty relations}

We shall continue the above discussion introducing the notion of
thermodynamic uncertainty relations and proposing their
generalization with the help of nonextensive statistics
\cite{GTR}. They were discussed in \cite{BH} where it was
suggested that the temperature $T$ and energy $U$ could be
regarded as being complementary in the same way as energy and time
are in quantum mechanics. A simple dimensional analysis suggests
that $ \Delta U\, \Delta \beta \ge k $, where $\beta = 1/T$ and
$k$ is Boltzmann's constant. Isolation ($U$ definite) and contact
with a heat bath ($T$ definite) are then the two extreme cases of
such complementarity. This is known as Lindhard's uncertainty
relation between the fluctuations of $U$ and $T$ \cite{JL}:
\begin{equation}
\omega_U^2\, +\, \omega^2_T\, =\, \frac{1}{\langle
N\rangle}\qquad{\rm where} \qquad  \omega^2_x = Var(x)/\langle
x\rangle^2 . \label{eq:JL}
\end{equation}
This idea is still disputable \cite{UL,UL1,UL2}, nevertheless we
can treat these increments as a measure of fluctuations of the
corresponding physical quantities. This allows us to analyze an
ensemble in which the energy ($U$), temperature ($T$) and
multiplicity ($N$), can all fluctuate and thus to express these
fluctuations by the corresponding parameters $q$. In this way,
using generalized thermodynamics (based on nonextensive
statistics) one gets the following relation \cite{GTR}
\begin{equation}
\Big| \omega^2_N - \frac{1}{\langle N\rangle}\Big| = \omega^2_U +
\omega^2_T - 2\rho \omega_U \omega_T = \left( \omega_U -
\omega_T\right)^2 + 2 \omega_U\omega_T(1 - \rho) = |q - 1|,
\label{eq:corq}
\end{equation}
where $\rho = \rho (U,T) \in [-1,1]$ is the correlation
coefficient between $U$ and $T$. This generalizes Linhard's
thermodynamic uncertainty relation.
\begin{figure}[tp]
\begin{center}
\includegraphics [width=11.5cm]{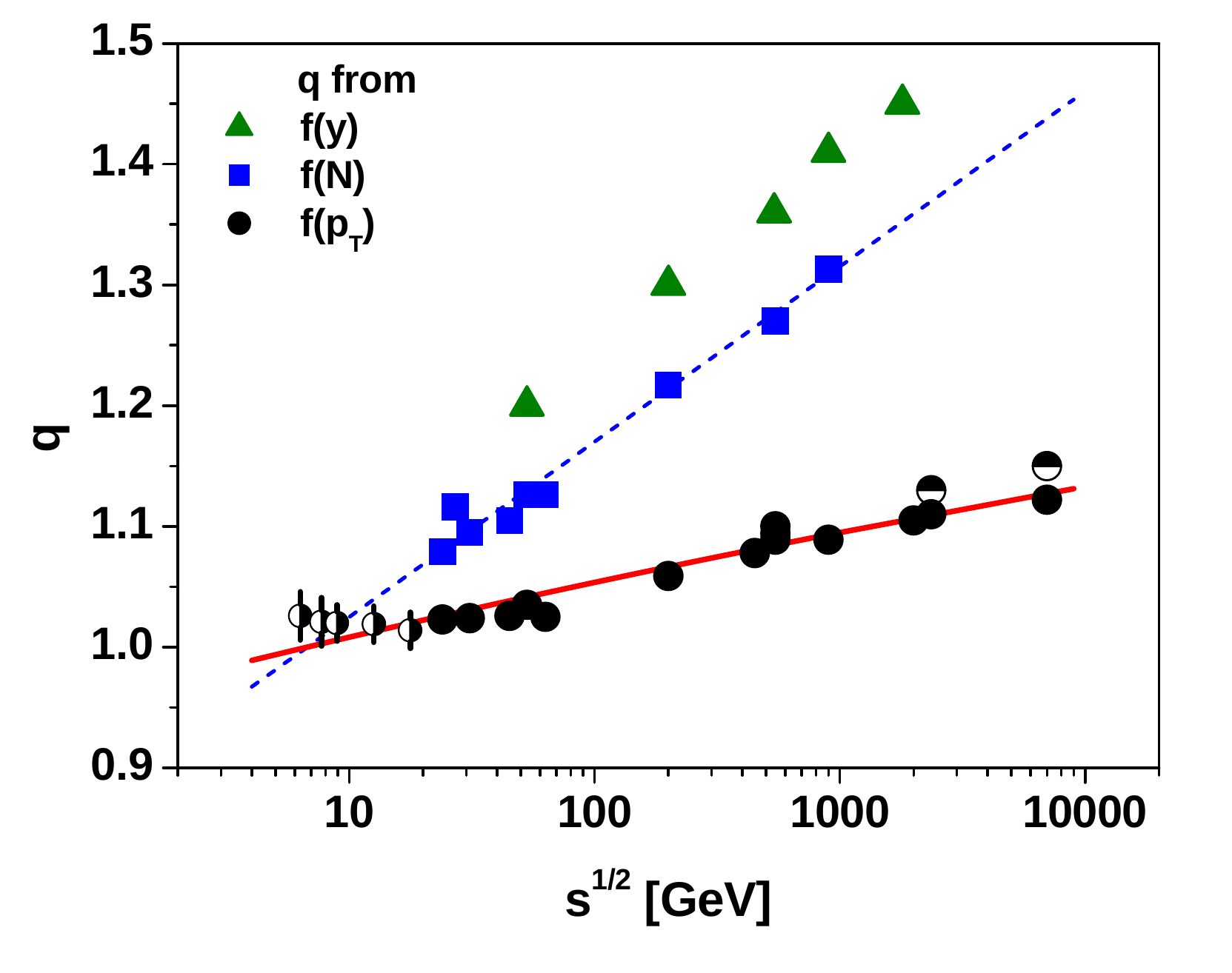}
\caption{(Color online) Energy dependencies of the parameters $q$
obtained from different observables. Triangles: $q$ obtained from
an analysis of rapidity distributions \cite{QlQt1,rapidity}.
Squares: $q$ obtained from multiplicity distributions $P(N)$
\cite{mult,mult1} (fitted by $ q = 1 + 1/k$ with $1/k = -0.104 +
0.029 \ln s$). Circles: $q$ obtained from different analysis of
transverse momenta distribution $f(p_T)$. Data points in this case
come, respectively, from the compilation of $p+p$ data (full
symbols) \cite{Wibig},  from CMS data (half filled circles at high
energies) \cite{LHC_CMS,LHC_CMS1} and from NA49 data from Pb+Pb
collisions (half filled circles at low energies)
\cite{NA49,NA49-1,NA49-2}. The full line comes from our
recalculation using Eq.(\ref{eq:corq}).} \label{FigA}
\end{center}
\end{figure}

The observed systematics in energy dependence of the parameter $q$
deduced from presently available data is shown in Fig. \ref{FigA}.
From measurements of different observables one observes that for
high enough energies $ q > 1$ (for low energies conservation laws
are important and one can encounter $ q < 1$ situations) and that
values of $q$ found from different observables are different. The
later is caused either by technical (methodical) problems or else
from some physical cause. The former arises when, for example,
fluctuations of temperature are deduced either from data averaged
over other fluctuations or from more refined data also accounting
for fluctuations of other variables (as in \cite{WWW}, see Eq.
(\ref{eq:sumrule})). The latter case is connected with the fact
that the observed $q$'s were obtained in different parts of phase
space (or both). In this case one gets an uncertainty relation
(\ref{eq:corq}) with the help of which one can connect
fluctuations observed in different parts of phase space. For
example, one can recalculate $q$ obtained from $P(N)$ (dashed line
in Fig. \ref{FigA}) to $q$ which can be evaluated from $f(p_T)$
(full line in Fig. \ref{FigA}), see \cite{GTR} for
details\footnote{A comment is necessary when looking at results at
Fig. \ref{FigA} obtained from $f(y)$. Namely, as observed in
\cite{qcompilation,WWW},  it turns out that in the fitting
procedure parameters $T$ and $q$ are strongly correlated. This is
why $q$ values evaluated in different analysis of rapidity
distributions \cite{QlQt1,rapidity} differ slightly from presented
here (roughly speaking, they give $q$ values comparable or
something higher that one obtained from multiplicity
distribution).}.

\section{\label{section:IV}Nonadditivity in nuclear collisions
(on the $q$ duality in nonextensive statistics)}

So far we were using Tsallis statistics without really resorting
to Tsallis entropy, i.e., we treated it as a kind of
superstatistics \cite{sS,sS1}. However, closer inspection of both
approaches reveals that corresponding nonextensivity parameters
(say $q$ and $q'$, respectively) are not the same, in fact one
encounters a sort of duality, like $q = 2 - q'$ discussed, for
example, in \cite{KGG,BJ,BJ1}. We shall now address this problem
in more detail on the example of nonadditivity observed in nuclear
collisions where, as we shall see, both types of $q$ can be
discussed and (in principle) compared at the same time \footnote{
Similar duality occurs in nonextensive treatment of fermions for
which the particle-hole correspondence, $n_q(E, T, \mu) = 1 -
n_{2-q}(-E, T, -\mu)$ (where $\mu$ is the chemical potential),
must be preserved by the q-Fermi distributions \cite{RW1,RW2}.
However, here we deal with different problem, namely that
parameter $q$ in entropy $S_q$ differs from parameter $q'$ in
probability distribution $f_{q'}$ and that $q= 2 - q'$.}.

One of the phenomenological approaches used to describe these
collisions is based on superposition models in which the main
ingredients are nucleons which have interacted at least once
\cite{WNM}. In this case, when sources are identical and
independent of each other, the total ($N$) and the mean ($\langle
N\rangle$) multiplicities are supposed to be given by,
\begin{equation}
N = \sum_{i=1}^{\nu}n_i, \qquad{\rm and}\qquad \langle N\rangle =
\langle \nu\rangle \langle n_i\rangle, \label{eq:Nu}
\end{equation}
where $\nu$ denotes the number of sources and $n_i$ the
multiplicity of secondaries from the $i^{th}$ source. Albeit at
present nuclear collisions are mostly described by different kinds
of statistical models \cite{MG_rev}, which automatically account
for possible collective effects, nevertheless a surprisingly large
amount of data can still be described by assuming the above
superposition of independent nucleon-nucleon collisions (possibly
slightly modified) as the main mechanism of production of
secondaries and the question of the range of its validity is a
legitimate one \cite{FW,FW1}.

Using the notion of entropy, and considering $\nu$ independent
systems for which the corresponding individual probabilities are
combined as
\begin{equation}
p^{(\nu)}_q\left(x_1,\dots,x_{(\nu})\right) =
\prod_{k=1}^{\nu}p^{(1)}_q\left( x_k\right), \label{eq:prodnu}
\end{equation}
and assuming that all $p^{(1)}_q\left(x_k\right)$ are the same for
all $k$ (i.e., their corresponding entropies $S_q^{(1)}$ are
equal), one finds that\footnote{Notice that $\ln \left[ 1 + (1 -
q) S^{(\nu)}_q\right] = \nu \ln \left[ 1 + (1 - q)
S^{(1)}_q\right]$ and $S^{(\nu)}_q \stackrel{ q \rightarrow
1}{\longrightarrow} \nu \cdot S^{(1)}_1$. For $q < 1$ one has
$S^{(\nu)}_q/\nu \stackrel{\nu \rightarrow
\infty}{\longrightarrow} \infty$, i.e., entropy $S^{(\nu)}_q$ is
nonextensive. For $ q  > 1$ one has $S^{(\nu)}_q \ge 0$ only for
$q < 1 + 1/S^{(1)}_q$ and $S^{(\nu)}_q/\nu \stackrel{\nu
\rightarrow \infty}{\longrightarrow} 0$, i.e., entropy is
extensive, $0 \leq S^{(\nu)}_q/\nu \le S^{(1)}_q$.}
\begin{equation}
S_q^{(\nu)} = \sum^{\nu}_{k=1}\frac{\nu !}{(\nu - k)!k!}(1 - q)^{k
- 1}\left[ S^{(1)}_q\right]^k = \frac{\left[ 1 + (1 - q) S^{(1)}_q
\right]^{\nu} - 1}{1 - q}. \label{eq:Snu}
\end{equation}
In the following we put $\nu = N_W/2 = N_P$ ($N_W$ is the number
of wounded nucleons and $N_P$ is the number of participants from a
projectile). Assuming naively that the total entropy is
proportional to the mean number of produced particles,
\begin{equation}
S = \alpha \langle N\rangle , \label{eq:SN}
\end{equation}
one obtains the following relation between mean multiplicities in
$AA$ and $NN$ collisions,
\begin{equation}
\alpha \langle N\rangle_{AA} = \frac{\left[1 + (1 - q)
\alpha\langle N\rangle_{pp}\right]^{N_P} - 1}{1 - q}.
\label{eq:alphaN}
\end{equation}
At this point we stress the following important observation, so
far not discussed in detail. Namely, because (as shown in
\cite{WWprc}), $\langle N\rangle_{AA}$ increases nonlinearly with
$N_P$ and $\langle N\rangle_{AA} > N_P \cdot \langle
N\rangle_{pp}$, the nonextensivity parameter obtained here from
considering the corresponding entropies must be smaller than
unity, $q < 1$. On the other hand, all estimations od the
nonextensivity parameter (let us denote it by $q'$) discussed
before lead to $q' > 1$. This is an apparent {\it $q$ duality in
nonextensive statistics}, on which we shall concentrate in more
detail.

Start with the obvious remark that, strictly speaking, relation
(\ref{eq:alphaN}) is not exactly correct for $S_q$. In what
follows we denote entropy on the level of particle production by
$s$ (and the corresponding nonextensivity parameter by
$\tilde{q}$), whereas the corresponding entropies and
nonextensivity parameter on the level of $NN$ collisions by $S$
and $q$. From Eq. (\ref{eq:Snu}) we have that for $N$ particles
\begin{equation}
s^{(N)}_{\tilde{q}} = \frac{\left[ 1 + \left(1  -
\tilde{q}\right)s^{(1)}_{\tilde{q}} \right]^N - 1}{1 - \tilde{q}}
\stackrel{\tilde{q} \rightarrow 1}{\longrightarrow} N\cdot
s^{(1)}_{\tilde{q}} = \alpha N, \label{eq:tildeq}
\end{equation}
where $s^{(1)}_{\tilde{q}} = \alpha$ is the entropy of a single
particle. In a $A+A$ collision with $\nu$ nucleons participating,
Eq. (\ref{eq:Snu}) results in
\begin{equation}
S^{(\nu)}_{q} = \frac{\left[ 1 + \left(1  - q\right)S^{(1)}_{q}
\right]^{\nu} - 1}{1 - q}, \label{eq:justq}
\end{equation}
where $S^{(1)}_q$ is the entropy of a single nucleon.

Denoting multiplicity in single $N+N$ collision by $n$, the
respective entropy is S$^{(1)}_q = S^{(1)}_{\tilde{q}} = \left\{
\left[ 1 + \left(1 - \tilde{q}\right)s^{(1)}_{\tilde{q}} \right]^n
- 1\right\}/(1 - \tilde{q})$, whereas entropy in $A+A$ collision
for $N$ produced particles is S$^{(N)}_{\tilde{q}} = \left\{
\left[ 1 + \left(1 - \tilde{q}\right)s^{(1)}_{\tilde{q}} \right]^N
- 1\right\}/(1 - \tilde{q})$. This means that
\begin{equation}
S^{(N)}_{\tilde{q}} = S^{(\nu)}_q. \label{eq:SS}
\end{equation}
\begin{figure}[tp]
\begin{center}
\includegraphics [width=11.5cm]{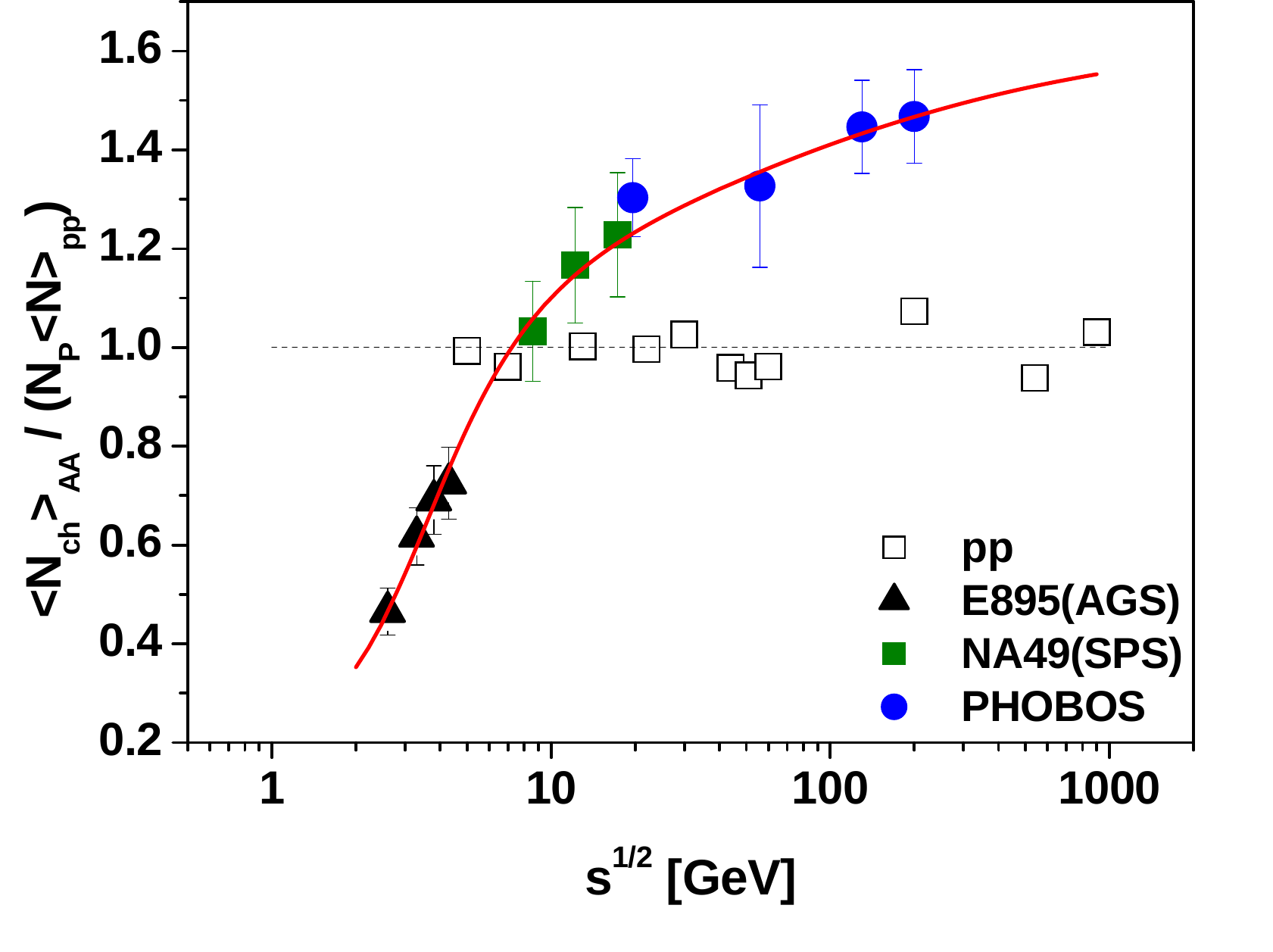}
\caption{(Color online)Energy dependence of the charged
multiplicity for nucleus-nucleus collisions divided by the
superposition of multiplicities from proton-proton collisions (cf.
Eq. (\ref{eq:FB})). Experimental data on multiplicity are taken
from compilation \cite{PHOBOS}.} \label{FigB}
\end{center}
\end{figure}
Notice that parameters $q$ and $\tilde{q}$ are usually not
identical. Moreover, from relation (\ref{eq:qN_P}) one gets that
for $NN$ collisions (where $N_P=A$) $\tilde{q} = 1$. On the other
hand, for $\tilde{q} = q$ Eq. (\ref{eq:SS}) corresponds to the
situation encountered in superpositions, as in this case one gets
that
\begin{equation}
\left[ 1 + (1 - q)s^{(1)}_q\right]^N = \left[1 + (1 -
q)s^{(1)}_q\right]^{n\nu}\qquad  {\rm or} \qquad N= n\nu.
\label{eq:Nnun}
\end{equation}

\begin{figure}[tp]
\begin{center}
\includegraphics [width=11.5cm]{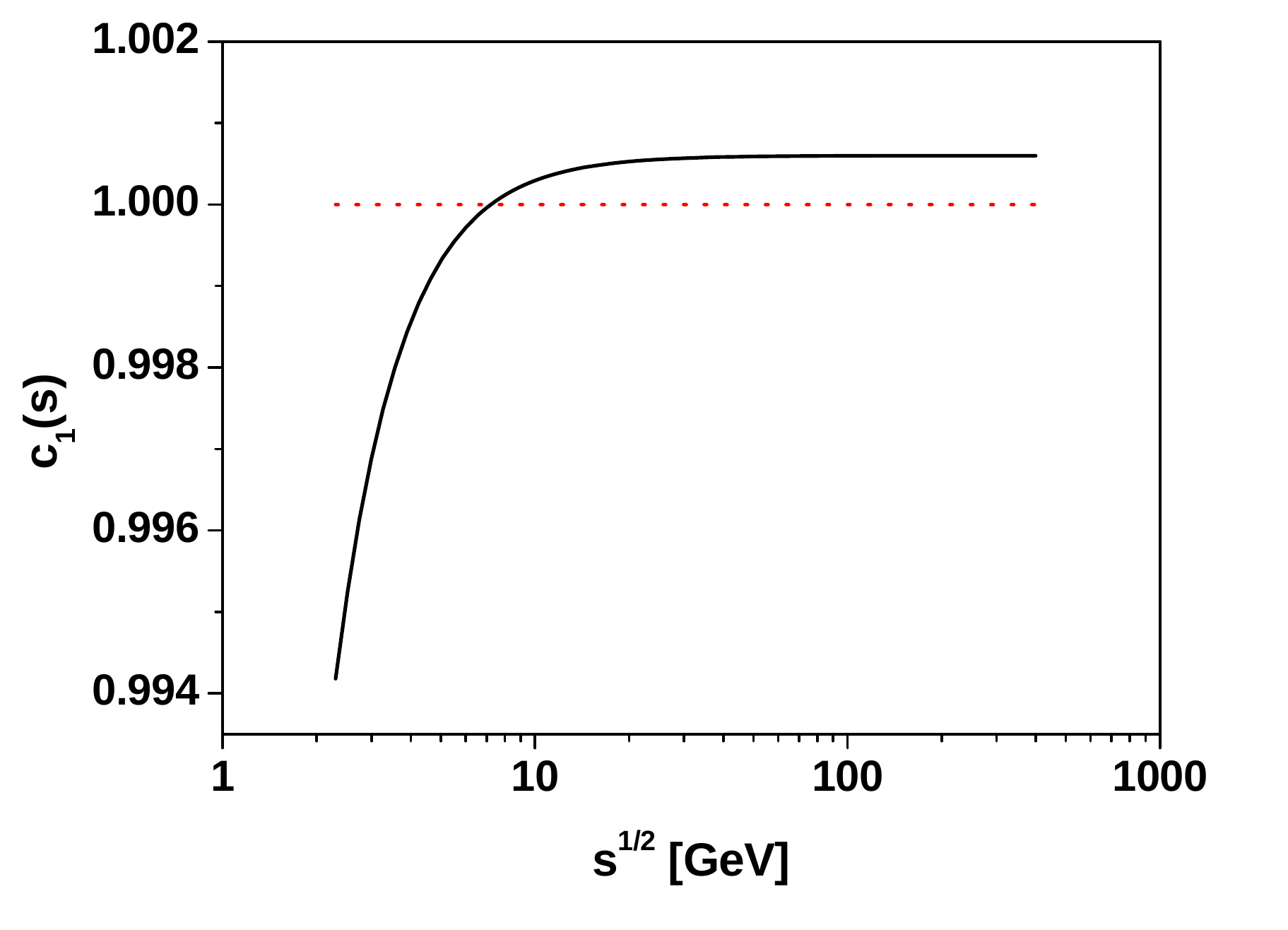}
\caption{(Color online)Energy ($\sqrt{s}$) dependence of the
parameter $c_1(s)$ obtained from the analysis of available data.}
\label{FigC}
\end{center}
\end{figure}

Consider now the general case and denote
\begin{equation}
c_1 = 1 + \left(1 -
\tilde{q}\right)s^{(1)}_{\tilde{q}};\qquad\qquad c_2 =
\frac{1-q}{1 - \tilde{q}}. \label{eq:c1c2}
\end{equation}
These quantities are not independent because:
\begin{equation}
c_2 c_1^N + 1 - c_2 = \left( c_2 c_1^n + 1 -
c_2\right)^{\nu}.\label{eq:ccc}
\end{equation}
From relation (\ref{eq:ccc}) one gets that
\begin{equation}
\frac{N}{\nu\cdot n} = \frac{1}{\nu n\cdot \ln c_1} \ln\left[
\frac{\left(c_2c_1^n + 1 - c_2\right)^{\nu} - \left( 1 -
c_2\right)}{c_2}\right], \label{eq:FB}
\end{equation}
which for $N = \langle N_{AA}\rangle $, $n = \langle
N_{pp}\rangle$ and $\nu = N_P$ is presented in Fig. \ref{FigB} for
different reactions. As seen there one can describe experimental
data by using $c_2 = 1.7$ and with $c_1$ depending on energy
$\sqrt{s}$ according to $c_1(s) = 1.0006 - 0.036 s^{-1.035}$, as
seen in Fig. \ref{FigC}. Notice that for energies $\sqrt{s} > 7$
GeV one has $c_1 > 1$. This means that $\tilde{q} < 1$ and
(because $c_2 > 0$) also $ q < 1$.

Now look at this problem from the view point of Tsallis entropy,
\begin{equation}
S_q = \frac{1}{1 - q}\left[ \int dx f^q(x) - 1 \right].
\label{eq:T}
\end{equation}
To get from it the probability density function $f(x)$, one either
optimizes it with constrains
\begin{equation}
\int dx f(x) = 1;\qquad  \int dx x f^q(x) = \langle x\rangle_q
  \label{eq:xq}
\end{equation}
and obtains \cite{RS}
\begin{equation}
f(x) = (2 - q)\left[ 1 - (1 - q)x\right]^{\frac{1}{1-q}};\qquad 0
\leq x < \infty;\quad 1\leq q\leq 3/2, \label{eq:T1}
\end{equation}
or else one uses as constrains
\begin{equation}
\int dx f(x) = 1;\qquad  \int dx x f(x) = \langle x\rangle
\label{eq:x}
\end{equation}
and obtains \cite{RS}
\begin{equation}
f(x) = \frac{q}{\left[ 1 + (1 - q)
x\right]^{\frac{1}{1-q}}};\qquad 0 \leq x < \infty;\quad 1/2 <
q\leq 1. \label{eq:T2}
\end{equation}
Notice now that only (\ref{eq:T1}) is the same as distribution
obtained in superstatistics and used above, cf., Eq.
(\ref{eq:Tsallis}). The second distribution, Eq. (\ref{eq:T2}),
which seems to be more natural from the point of view of a
physical interpretation of the constraint used, becomes the first
one if expressed in $q'$ given by
\begin{equation}
q' = 2 - q , \label{eq:equivalence}
\end{equation}
namely, in this case one has
\begin{equation}
f(x) = (2 - q')\left[ 1 - (1 - q')x\right]^{\frac{1}{1-q'}}.
\label{eq:qprime}
\end{equation}
We show here, cf. Fig. \ref{FigA}, that using a Tsallis
distribution in the form of Eq. (\ref{eq:qprime}), one gets $q' >
1$. On the other hand, non additivity in the superposition model
described using the notion of entropy clearly requires $q < 1$,
cf. Figs. \ref{FigB} and \ref{FigC}. This means that $q'$ is not
the same as $q$. The conclusion one can derive from these
considerations is that the second way of deriving $f(x)$, which
uses a linear condition, cf. Eq. (\ref{eq:T2}), is the correct one
and that $q'$ in distribution is not the same as $q$ in entropy.
The problem is that, whereas from distributions one can easily
deduce a numerical value of $q'$, this is not the case when one
uses entropy. There are too many variables to play with (cf.,
considerations using the superposition model as above). For
example, in the definition of $c_1$ in Eq. (\ref{eq:c1c2}), one
has the $s^{(1)}_{\tilde q}$, which is not known {\it a priori}.
The only thing one can get in this case is that $q < 1$. We cannot
therefore check numerically that relation (\ref{eq:equivalence})
really holds. But, if one agrees that the Tsallis distribution
comes from Tsallis entropy, we have only two options: either $q' =
q$ or $q' -1 = 1 -q$. Our conclusion presented here, that $q' > 1$
and $q < 1$, therefore supports the second option, i.e., Eq.
(\ref{eq:equivalence}).

\section{\label{section:V}Summary}

To summarize, Tsallis statistics is fruitful because in a very
economical way (with only one new parameter $q$) it describes the
power-like behavior of different observables. This parameter, for
$q > 1$ considered here, is given fully by nonstatistical
fluctuations present in the system and visible as fluctuations of
the scale parameter in superstatistics. It also allows (via
specific sum rules or through a generalized thermodynamic
uncertainty relation) to connect fluctuations of different
observables or observed in different parts of phase space.
Finally, when considering a superposition scenario, for example,
in the scattering of nuclei, the relation $q' - 1 = 1 - q$ seems
to be observed (with $q'$ occurring in the Tsallis distribution
and $q$ in Tsallis entropy).\\

This final observation needs some more attention. The probability
density function (PDF) is commonly evaluated by the Maximum
Entropy Method (MEM) for Tsallis entropy with some constraints
\cite{OM} \footnote{Notice that Tsallis entropy is a monotonic
function of the Renyi entropy, $S_q =
\ln_q\left[\exp\left(R_q\right)\right]$, and that both lead to the
same equilibrium statistics of particles (with coinciding maxima
in equilibrium for similar constraints on the expectation
value).}. At the moment, there are four possible MEMs discussed at
length in \cite{E6} using two kinds of definition for an
expectation value of physical quantities: the normal average
(\ref{eq:x}) and the $q$-average (\ref{eq:xq}) (with normal, as
here, or the so-called escort PDFs \cite{escort,escort2,escort1}).
Various arguments have been given justifying the $q$-average
\cite{E11,E12,E13}. Recently, however, it has been pointed out
that, for a small change of the PDF, thermodynamic averages
obtained by the $q$-averages are unstable, whereas those obtained
by the normal average are stable \cite{E14,E15}. On the other
hand, it is claimed \cite{E17} that for the escort PDF, the
Tsallis entropy and thermodynamical averages are robust. This
means that this issue on the stability (robustness) of
thermodynamical averages as well as the Tsallis entropy is still
controversial \cite{E18}.

\section*{Acknowledgements}
Partial support (GW) of the Ministry of Science and Higher
Education under contract DPN/N97/CERN/2009 is acknowledged. We
would like to thank warmly dr Eryk Infeld for reading this
manuscript.

\end{document}